\documentclass[prb,aps,amsmath,amssymb,notitlepage,twocolumn,floatfix,superscriptaddress]{revtex4-1}
\usepackage{amsthm}
\usepackage{amsfonts}
\usepackage{siunitx}
\usepackage{amsmath}
\usepackage{amssymb}
\usepackage{graphicx}
\usepackage{verbatim}
\usepackage[colorlinks]{hyperref}
\usepackage{tikz}
\usepackage{pgfplots}
\usepackage{braket}

\definecolor{linkcolor}{RGB}{0,83,166}
\hypersetup{
    colorlinks = true,
    allcolors = {linkcolor}
}

\pgfplotsset{compat=newest}
\usetikzlibrary{plotmarks}
\usepgfplotslibrary{groupplots}

\pgfdeclarelayer{foreground}
\pgfsetlayers{main,foreground}

\newlength\figureheight
\newlength\figurewidth

\renewcommand{\epsilon}{\varepsilon}

\pgfplotsset{every axis/.append style={
    axis line style=thick,
  }
}

\usetikzlibrary{external}
\begin{document}
\title{Experimental demonstration of perturbative anticrossing mitigation using non-uniform driver Hamiltonians}
\author{Trevor Lanting}\email[]{tlanting@dwavesys.com}
\affiliation{D-Wave Systems Inc., Burnaby B.C.}
\author{Andrew D.~King}\email[]{aking@dwavesys.com}
\affiliation{D-Wave Systems Inc., Burnaby B.C.}
\author{Bram Evert}\email[]{bevert@dwavesys.com}
\affiliation{D-Wave Systems Inc., Burnaby B.C.}
\author{Emile Hoskinson}\email[]{ehoskinson@dwavesys.com}
\affiliation{D-Wave Systems Inc., Burnaby B.C.}
\date{\today}

\begin{abstract}
Perturbative anticrossings have long been identified as a potential computational bottleneck for quantum annealing.
This bottleneck can appear, for example, when a uniform transverse driver Hamiltonian is applied to each qubit.
Previous theoretical research sought to alleviate such anticrossings by adjusting the transverse driver Hamiltonians on individual qubits according to a perturbative approximation.
Here we apply this principle to a physical implementation of quantum annealing in a D-Wave 2000Q system.
We use samples from the quantum annealing hardware and per-qubit anneal offsets to produce nonuniform driver Hamiltonians.
On small instances with severe perturbative anticrossings, our algorithm yields an increase in minimum eigengaps, ground state success probabilities, and escape rates from metastable valleys.
We also demonstrate that the same approach can mitigate biased sampling of degenerate ground states.
\end{abstract}

\maketitle


\section{Introduction}

Quantum annealing (QA) holds the potential to confer a computational advantage over classical methods via multiqubit tunneling and entanglement \cite{Lanting2014, Denchev2016}.  In the near term it offers a scalable alternative to circuit-model quantum computing \cite{Andriyash2017, Denchev2016}.  QA is based on physical evolution of a time-dependent quantum Hamiltonian towards a classical Hamiltonian; the fact that the mid-anneal Hamiltonian may not resemble the final Hamiltonian provides both an advantage and a challenge.  In some cases, QA fails with high probability when the instantaneous quantum ground state has overwhelming support from excited states in the computational basis.  As the anneal progresses this support can vanish at a small-gap perturbative anticrossing, where QA fails via Landau-Zener transition.

Previously available quantum annealing systems have allowed only uniform QA, in which each qubit is initialized with the same transverse-field driver Hamiltonian and all qubits are annealed in unison.  In this case small-gap perturbative anticrossings can arise in unfavorably structured inputs, even those that are not particularly hard for classical solvers \cite{Boixo2013,Steiger2015,tailspaper}.

Just as this computational bottleneck can be identified via a perturbative expansion, it can be mitigated with perturbative expansion.  Such approaches, in particular those that modify the transverse field on a per-qubit basis, have been proposed and simulated \cite{Dickson2012}.  Here we experimentally demonstrate a similar approach in which qubit dynamics are tuned via {\em anneal offsets}, which allow for suppression or enhancement of a qubit's dynamics by annealing the qubit slightly in advance of or behind other qubits, respectively.  We find that both the longitudinal and the transverse perturbative corrections contribute to improved performance across a set of tailored inputs.  Measurements of tunneling dynamics and spectral analysis support these findings.

\section{Mitigating perturbative anticrossings with non-uniform driver Hamiltonians}

\begin{figure}
  \includegraphics{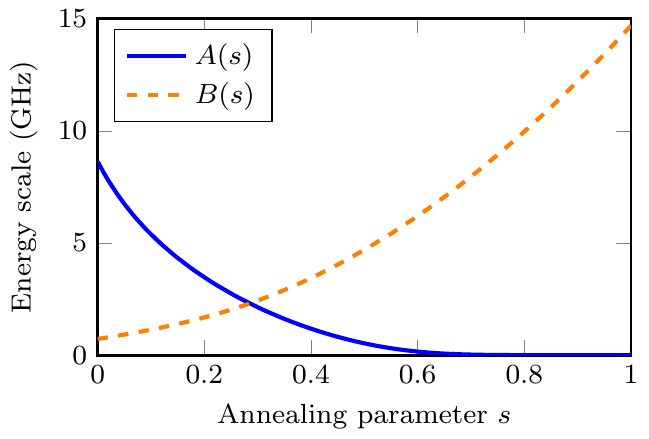}
  \caption{\label{FIG:uniform-annealing}  $A(s)$ and $B(s)$ versus $s$ for the QA system under the uniform QA algorithm.  A global time-dependent annealing bias tunes both $A(s)$ and $B(s)$ simultaneously for all qubits throughout the QA algorithm.}
\end{figure}

We consider a system of qubits connected to one another with tunable longitudinal spin-spin interactions $J_{ij}$.  We can write the uniform QA Hamiltonian as a time-dependent linear combination of an initial driver Hamiltonian $\mathcal H_D$ and a final problem Hamiltonian $\mathcal H_P$
\begin{eqnarray}
\mathcal H(s) &=& \tfrac{1}{2}A(s)\mathcal H_D + \tfrac{1}{2}B(s)\mathcal H_P\label{eqn:uniform}\\
\mathcal H_D &=& -\sum_i \sigma_x^i\label{eqn:driver}\\
\mathcal H_P &=& \sum_i h_i\sigma_z^i + \sum_{i<j}J_{ij}\sigma_z^i\sigma_z^j\label{eqn:problem}
\end{eqnarray}
where $A(s)$ is the transverse driver energy, $B(s)$ is the problem energy, $s = t/t_f$ is a normalized annealing parameter, $\sigma_x^i, \sigma_z^i$ are Pauli matrices operating on the $i$th qubit. At the beginning of the QA algorithm, $A(0) \gg B(0)$ and at the end of the annealing algorithm $A(1) \ll B(1)$.  Fig.~\ref{FIG:uniform-annealing} shows the annealing schedule $\{A(s),B(s)\}$ for the system used herein.  All instances studied have $h_i=0$ for each qubit.

\subsection{Generalizing the uniform QA algorithm}

The terms $A(s)$ and $B(s)$ represent the transverse and longitudinal energy scales, which in the uniform QA algorithm do not differ from qubit to qubit.  Allowing non-uniform energy scales $A_i(s)$ and $B_i(s)$ on qubit $i$, we can rewrite $\mathcal H$:
\begin{multline}
\mathcal H(s) = -\tfrac{1}{2}\sum_i A_i(s)\sigma^i_x + \tfrac{1}{2}\sum_i B_i(s)h_i\sigma^i_z\\+ \tfrac{1}{2}\sum_{i<j}\sqrt{B_i(s)B_j(s)}J_{ij}\sigma^i_z\sigma^j_z.\label{eq:generalized}
\end{multline}

\subsection{Connecting perturbative anticrossings and degeneracy}\label{sec:perturbation}

A spin in a classical state is said to be {\em free} or {\em floppy} if flipping the spin results in another classical state that is degenerate.  Floppy qubits, and associated degeneracies, have been implicated in perturbative anticrossings \cite{tailspaper, boixo2014evidence, Boixo2013}.  Here we provide an argument, based on first-order perturbation calculations, to connect observed probabilities of qubit floppiness to the change in energy of degenerate excited states when a finite transverse field is present. This analysis, from which we derive our mitigation strategy, closely mirrors the approach of Dickson and Amin in Ref.~\cite{Dickson2012}, where the authors use a second-order perturbative expansion because the first order is zero in the problem studied.

We first consider a system with $n$ qubits with two-fold degeneracy in the second classical energy level. We can label these excited classical states $\ket{\alpha}$ and $\ket{\beta}$, each having energy $E_1$. We also suppose that $\ket{\alpha}$ and $\ket{\beta}$ differ by a single bit flip. At $s = 1-\epsilon$, a small transverse field $A(s)$ is present which lifts the degeneracy between these states. The new eigenenergies are given by

\begin{equation} E_1' = E_1 + \lambda_k \end{equation}
where $\lambda_k$ are the two eigenvalues of the matrix

\begin{eqnarray}
V &=& \tfrac{1}{2}A(s)
\begin{bmatrix}
    \braket{\alpha|\mathcal H_D|\alpha} & \braket{\alpha|\mathcal H_D|\beta} \\
    \braket{\beta|\mathcal H_D|\alpha} & \braket{\beta|\mathcal H_D|\beta} \\
\end{bmatrix}\\
&=& \tfrac{1}{2}A(s)
\begin{bmatrix}
    0 & -1 \\
    -1 & 0 \\
\end{bmatrix}.
\end{eqnarray}

In this simple example, $\lambda_0 = -A(s)/2$ and $\lambda_1 = A(s)/2$. To first order, the presence of $A(s)$ lifts the degeneracy, lowering the energy of the symmetric superposition of $\ket{\alpha}$ and $\ket{\beta}$ by $A(s)/2$. Since $A(s)$ grows as $s \rightarrow 0$, this produces a perturbative anticrossing between the ground state and the first excited state.

We now consider a system with $n$ qubits and $N$ degenerate excited states with energy $E_1$. When these states are connected via single bit flips, the presence of a transverse field $A(s)$ again lifts the degenerate of these $N$ states. To first order, the lowest energy

\begin{equation}
  E_1' = E_1 - \Delta'
  \end{equation}
and
\begin{equation}
  \Delta' = \frac{1}{N}\sum_{a=1}^N\sum_{b:(a,b)\in B} \tfrac{1}{2}A(s),
  \end{equation}
where $B$ is the set of pairs of states $a, b$ connected by a single bit flip (see Appendix~\ref{app:perturbation} and \cite{amin-choi}). Defining $q(a,b)$ as the qubit index associated with the single bit flip difference, we can write
\begin{equation}
  \Delta' = \tfrac 12 A(s) \sum_{i=1}^n \frac{1}{N}\sum_{a=1}^N\left( \sum_{b:(a,b)\in B}\delta(q(a,b)-i)\right)
  \end{equation}
or
\begin{equation}
  \Delta' = \tfrac 12 A(s) \sum_i^n \left< \sum_{b:(a,b)\in B}\delta(q(a,b)-i)\right>_N,
  \end{equation}
where $\delta$ denotes the Dirac delta function.  We define a per-qubit floppiness metric
\begin{equation}
  F_i = \left< \sum_{b:(a,b)\in B}\delta(q(a,b)-i)\right>_N,
  \end{equation}
which can be interpreted as the fraction of the $N$ states in which the $i$th qubit is floppy. To first order, then, there is a relationship between qubit floppiness and the reduction in excited state energy:

\begin{equation}
  \Delta' = \tfrac 12 \sum_i^n A(s)F_i.
\end{equation}
Again, $A(s)$ grows as $s\rightarrow 0$, producing a perturbative anticrossing between the ground state and the first excited state.

This suggests a simple mitigation algorithm based on producing a non-uniform per-qubit driver Hamiltonian $A_i(s)\sigma_x^i$:  If an initial set of samples obtained via the ``uniform'' annealing algorithm is dominated by states from a highly degenerate cluster of first excited states, we can reduce $\Delta'$ by decreasing $A_i(s)$ for qubits for which $F_i$ is high relative to other qubits, based on the empirical samples returned by the hardware. If there are multiple clusters of first excited states, this algorithm based on $F_i$ should also lift the lowest energy of each of these clusters. More generally, the goal is to suppress the appearance of states that are disproportionately favored by a perturbative transverse field, and simply mitigating based on these observed states will suppress oversampled states disproportionately.

Before describing our algorithmic implementation of this approach, we need to lay out the limitations of our {\em in situ} per-qubit anneal control.

\subsection{Per-qubit anneal offsets}

Currently available QA systems do not facilitate independent control of $A_i(s)$ and $B_i(s)$.  Instead we control $A_i(s)$ and $B_i(s)$ together on a per-qubit basis using {\em anneal offsets}.  

\begin{figure}
  \includegraphics{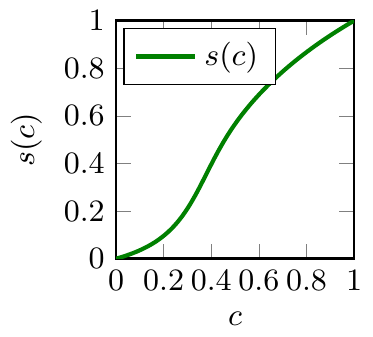}\\
  \includegraphics{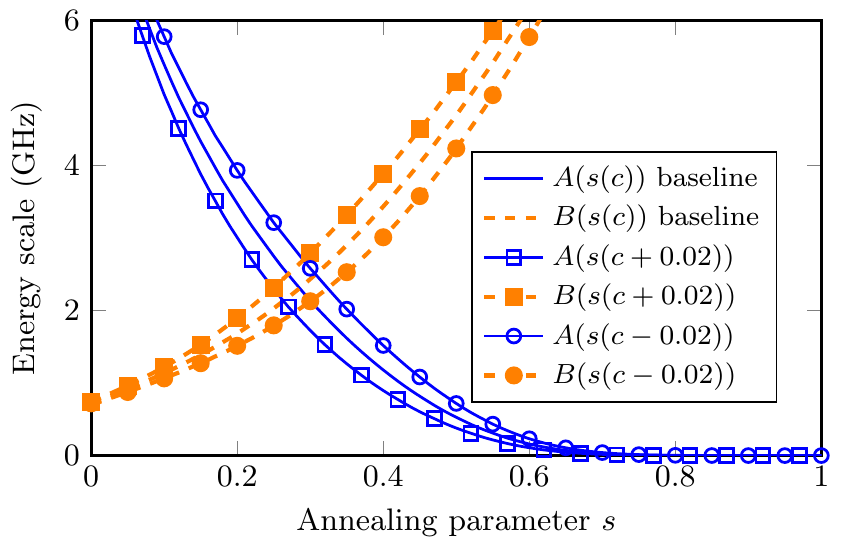}
  \caption{\label{FIG:offset-annealing} (Top) relationship between $s$ and the tunable parameber $c$.  (Bottom) $A(s)$ and $B(s)$ versus $s$ for the QA hardware and for $\delta_i = 0.0$, $0.02$, and $-0.02$.}
\end{figure}

The D-Wave 2000Q system by default implements Hamiltonian~(\ref{eqn:uniform}) with networks of coupled rf-SQUID flux qubits~\cite{Lanting2014}.  See Appendix~\ref{app:methods} for more details.  The QA algorithm is run by adjusting a global time-dependent bias, $\Phi^x_{\rm CCJJ}$, simultaneously for all of the qubits. We define a normalized global bias, $c$, as

\begin{equation}
  c = \frac{\Phi^x_{\rm CCJJ} - \Phi_{\rm CCJJ}^i}{\Phi_{\rm CCJJ}^f - \Phi_{\rm CCJJ}^i}
\end{equation}
where $\Phi^i_{\rm CCJJ}/\Phi_0 = -0.6457$ and $\Phi^f_{\rm CCJJ}/\Phi_0 = -0.7140$ for the particular processor used in this study. Like the annealing parameter $s$, $c$ ranges from $0$ at the beginning of the anneal to $1$ at the end of the anneal. Increasing $c$ over the course of the anneal both decreases $A(s)$ and increases $B(s)$ for all qubits; the ratio $A(s(c))/B(s(c))$ is a fixed function of $c$ determined by the macroscopic rf-SQUID flux qubit device parameters (see Appendix~\ref{app:methods}).  Fig.~\ref{FIG:uniform-annealing} shows $A(s(c))$ and $B(s(c))$ for the experiments discussed herein, and Fig.~\ref{FIG:offset-annealing} shows $s(c)$. Note that we choose this particular global trajectory to produce a quadratic growth of $B(s(c))$. 

We can modify the annealing trajectory for an individual qubit through the use of a tunable {\em in situ} static bias $\delta_i$ that adds or subtracts from the global time-dependent annealing bias signal $c$, giving each qubit a bias signal $c_i = c + \delta_i$.  We then define $A_i(s)$ as $A(s(c+\delta_i))$ and $B_i(s)$ as $B(s(c+\delta_i))$.  Fig.~\ref{FIG:offset-annealing} shows $A_i(s)$ and $B_i(s)$ for $\delta_i \in \{ -0.02, 0,0.02\}$. On current hardware, the tunable static bias allows an adjustment of up to $\delta_i \sim 0.1$. This allows one to advance or delay the annealing signal locally for each qubit.

\subsection{Algorithmic mitigation approach}\label{sec:algorithm}

Following the perturbative analysis in Section \ref{sec:perturbation}, we employ an iterative method similar to the one described in Ref.\ \cite{Dickson2012}.  At each iteration $k$, per-qubit anneal offsets $\delta_{i,k}$ are adjusted to slow the dynamics of qubits that are frequently floppy in the observed states.  Adjustments are scaled down for each successive iteration as the iterative search is refined by combining the new and current offsets in an ansatz ratio of $\sqrt k : 1$.

\begin{enumerate}
\item Choose a static offset magnitude $\alpha$.
\item Initialize each anneal offset $\delta_{i,0}$ to zero.
\item For iteration $k = 1,\ldots,n_{\mathit{iter}},$ 
  \begin{enumerate}
  \item Anneal $r$ times, saving each result.
  \item Compute $\mu_{i,k}$, the probability of floppiness, for each qubit based on the $r$ saved results of the current annealing run.
  \item Adjust each anneal offset according to $\mu_{i,k}$:
    $$\delta_{i,k} := \delta_{i,k-1} + \frac{\alpha\mu_{i,k} - \delta_{i,k-1}}{1+\sqrt k}.$$
  \end{enumerate}
\end{enumerate}

\begin{figure}
  \setlength{\figurewidth}{2.1cm}
  \setlength{\figureheight}{3cm}
  \includegraphics[width=6cm]{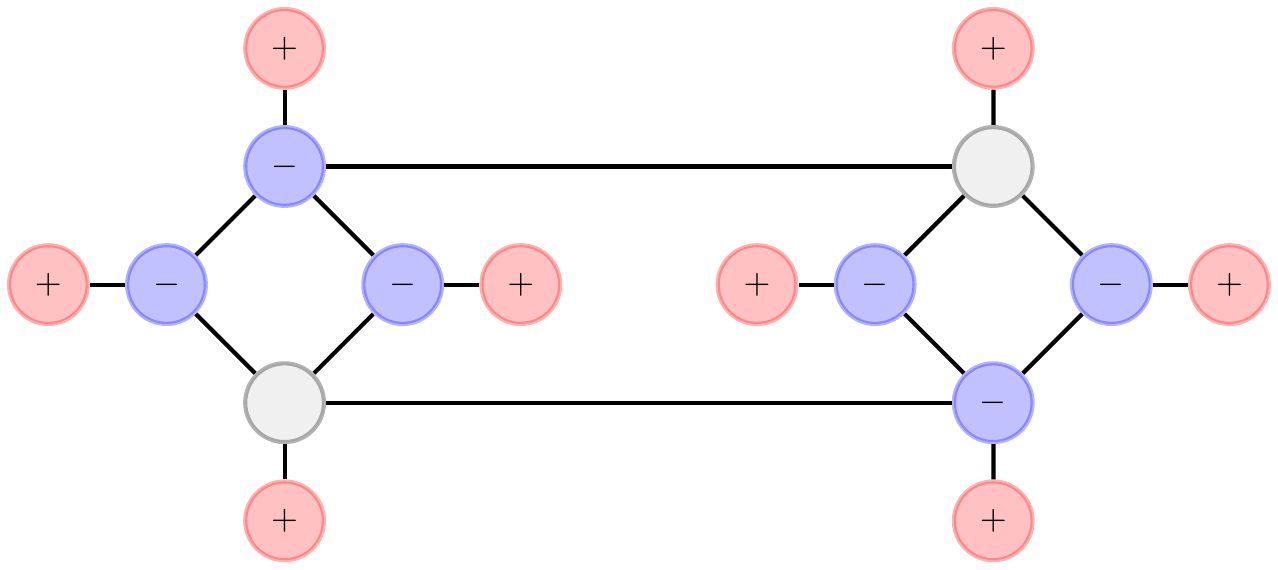}\\\vspace{.5cm}

  \includegraphics{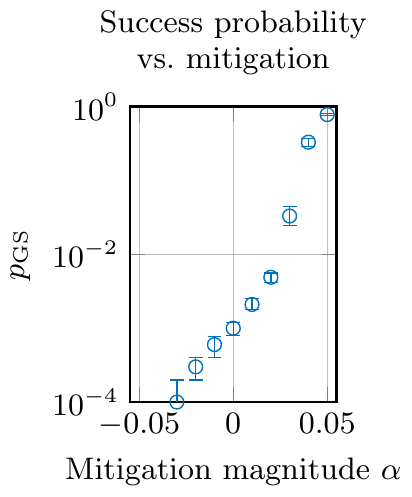}
  \includegraphics{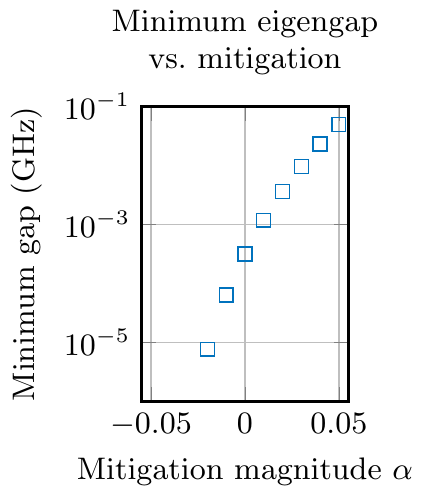}
  \caption{\label{fig:dickson} (Top) 16-qubit instance studied in Ref.~\cite{Dickson2013}; circles represent qubits with fields of value $+1$ (red), $0$ (gray), and $-1$ (blue), and lines represent FM couplings with value $-1$.  (Bottom) A single iteration of our mitigation strategy advances the outer qubits relative to the inner qubits.  Observed ground state probability (bottom left) increases with mitigation magnitude $\alpha$, as does the minimum eigengap (bottom right).}
\end{figure}

As a simple demonstration of the method we apply one iteration of the method to the 16-qubit system studied in Ref.~\cite{Dickson2013}, similar to those studied elsewhere \cite{Boixo2013,Albash2015}, shown in Fig.~\ref{fig:dickson}.  This instance has a unique ground state and a metastable valley of $2^8$ first excited states in which all eight outer qubits (those coupled to only one other qubit) are floppy, leading to a small-gap anticrossing and low ground state probability when run without mitigation.  In the first iteration if mitigation, the outer qubits have $\mu_{i,1} \approx 1$ and the others have $\mu_{i,1}\approx 0$.  Applying mitigation with $\alpha > 0$ increases both ground state probability and minimum eigengap, while the opposite is true when $\alpha < 0$.

\section{Improving optimization with anneal offsets}

\subsection{Optimization testbed}

\begin{figure}
  \includegraphics[width=3cm,trim={6mm 6mm 6mm 6mm},clip]{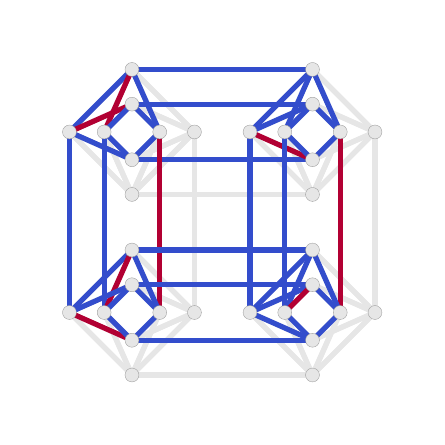}
\caption{A randomly-generated 24-qubit instance shown on a $2\times 2$ grid of unit cells in the D-Wave 2000Q qubit topology.  Qubits are represented by circles; couplers are represented by line segments.  Each nonzero coupler $J_{ij}$ is $+1$ (red) or $-1$ (blue).  This instance has two classical ground states $\uparrow\uparrow\cdots\uparrow$ and $\downarrow\downarrow\cdots\downarrow$, and $424$ first excited states.\label{fig:problem}}
\end{figure}

To study the effect of mitigating perturbative crossings with nonuniform driver Hamiltonians, we need small instances---small enough to exactly diagonalize Hamiltonian~(\ref{eqn:uniform})---in which the uniform QA algorithm will be drawn into a large valley of metastable states by perturbative anticrossings as described in Section \ref{sec:perturbation}.  We begin with a qubit connectivity graph with 24 qubits, with each qubit coupled to four others, and construct many thousands of random Ising instances by randomly assigning each $J_{ij}$ a value of $+1$ or $-1$.  Fig.~\ref{fig:problem} shows an example.
The median ground state probability observed on these instances using the uniform QA algorithm is over $99\%$.  To identify a set of instances for which uniform QA fails due to perturbative anticrossings, we first discard all instances with more than three ground states or fewer than 50 first excited states.  We then run all remaining instances with the uniform QA algorithm, performing $10^4$ anneals and collecting all spin configurations.  Our testbed consists of those 100 instances---unique up to graph isomorphism and spin reversal---for which the processor yielded the lowest observed ground state probability. Each instance has exactly two classical ground states forming a symmetric pair: $\uparrow\uparrow\cdots\uparrow$ and $\downarrow\downarrow\cdots\downarrow$.  Each has large valleys of first excited states connected by floppy qubits.

\subsection{Ground state probability}\label{sec:pgs}

\begin{figure}
\setlength{\figurewidth}{2.8cm}
\setlength{\figureheight}{2.8cm}
\includegraphics{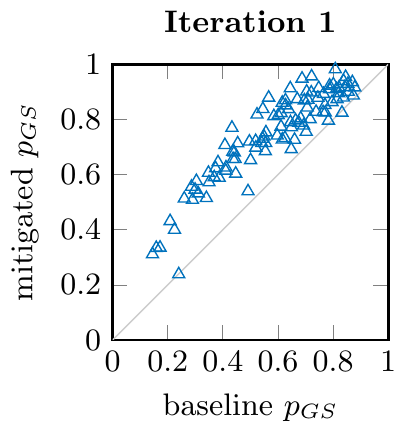} \includegraphics{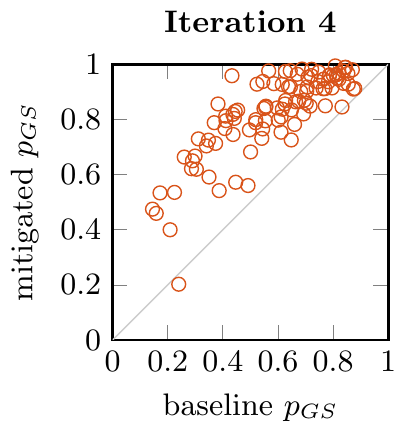}

\caption{\label{fig:pgs-results} Algorithmic mitigation improves QA success probability.  Shown are ground state probabilities $p_{\mathrm{GS}}$ after zero, one, and four iterations of mitigation.  Median $p_{\mathrm{GS}}$ increases from 62\% to 85\% over four iterations.}
\end{figure}

Fig.~\ref{fig:pgs-results}
 shows measurements of ground state probability $p_{\rm GS}$ for the 100 testbed instances after zero, one, and four iterations of the algorithm outlined in Section \ref{sec:algorithm} for all 100 instances. We run with $\alpha = 0.04$ and $r_k = 3.15\times 10^5$.  The median $p_{\rm GS}$ improves from $62\%$ to $85\%$ over four iterations of mitigation.

\begin{figure}
\setlength{\figurewidth}{4cm}
\setlength{\figureheight}{4cm}
\includegraphics{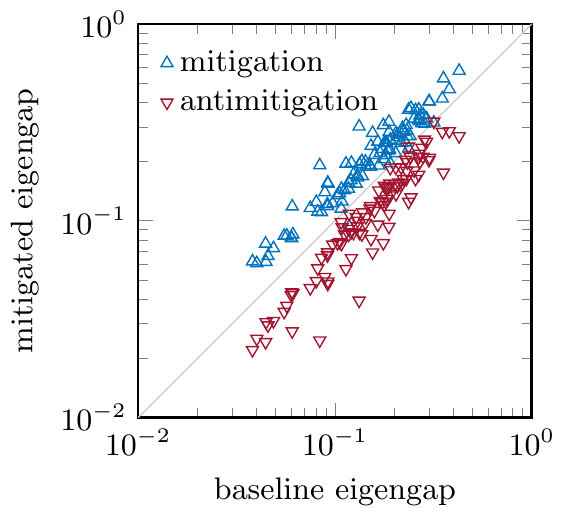}
\caption{Algorithmic mitigation increases the spectral gap.  Shown are the minimum gap between the instantaneous ground state and third excited state.\label{fig:eigengaps}}
\end{figure}

We examine the minimum eigengaps calculated for these instances---see Appendix \ref{app:eigengap}---and compare to the minimum eigengaps calculated after a single iteration of the mitigation algorithm applying offsets $\delta_i = \alpha\mu_i$ for both $\alpha = 0.02$ and $\alpha = -0.02$, which we term {\em mitigation} and {\em antimitigation} respectively. Fig.~\ref{fig:eigengaps} shows the mitigated and antimitigated eigengaps plotted versus the baseline eigengap. The algorithm systematically increases the minimum eigengap for $\alpha = 0.02$ and systematically decreases the eigengap for $\alpha = -0.02$.

\subsection{Approximating orthogonal control of $\mathcal H_D$ and $\mathcal H_P$}\label{sec:orthogonal}

Anneal offsets do not offer control of a qubit's transverse field independent of classical energy scale.  Advancing a qubit slows its dynamics by both reducing its transverse field and increasing its Ising energy scale.  To determine whether or not the positive effects of mitigation can be explained by variation of classical energy scale alone, we repeat the experiment from Section \ref{sec:pgs}, now compensating for variation in classical energy scale.

\begin{figure}
  \setlength{\figurewidth}{2.8cm}
  \setlength{\figureheight}{2.8cm}
\includegraphics{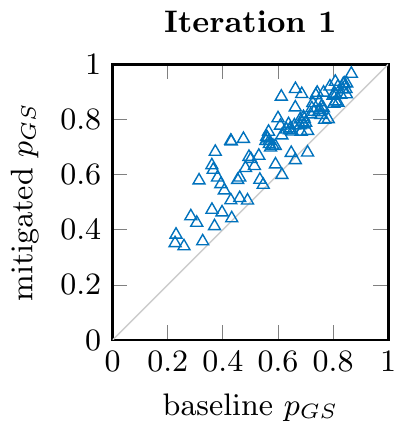} \includegraphics{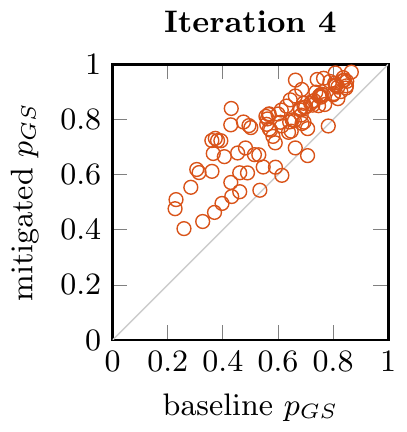}
\caption{\label{fig:orthogonal} Application of anneal offsets causes imbalance in longitudinal fields between qubits.  We approximate orthogonal transverse field control by compensating for this imbalance in the Ising Hamiltonian sent to the QA system.  Once this compensation is applied, mitigation still achieves a systematic improvement in success probability, albeit smaller than the improvement seen in Fig.~\ref{fig:pgs-results}.}
\end{figure}

To balance classical energy scales at a particular point $s^*$, we find a new set of couplings $J'_{ij}$ that equalize effective energy scales for each coupler at $s^*$:
$$
\sqrt{B_i(s^*)B_j(s^*)}J'_{ij} = B(s^*)J_{ij}.
$$
Following (\ref{eq:generalized}), applying QA to $J'$ in the place of $J$ gives us
\begin{equation}
\mathcal H(s^*) = -\frac{1}{2}\sum_i A_i(s^*)\sigma^i_x + \frac{1}{2}B(s^*)\sum_{i<j}J_{ij}\sigma^i_z\sigma^j_z,
\end{equation}
approximating independent control of $A_i$ at $s^*$.  We fix $s^*= 0.3$, where $J'_{ij}$ and $J_{ij}$ differ the most.  Results are shown in Fig.~\ref{fig:orthogonal} using $\alpha = 0.04$.

Mitigation continues to provide a systematic performance improvement, but less so than what was achieved without balancing classical energy scales.  This indicates that both factors contribute to improved performance in the $p_{\mathrm{GS}}$ metric.  When running these experiments we rescaled all Ising Hamiltonians by a factor of $0.85$ to keep all couplings within the available interval $[-2,1]$.  The increase in minimum baseline $p_{\mathrm{GS}}$ when downscaling the Hamiltonian is consistent with previous work \cite{Albash2015,tailspaper}.

\begin{figure}
  \setlength{\figurewidth}{4cm}
  \setlength{\figureheight}{4cm}
\includegraphics{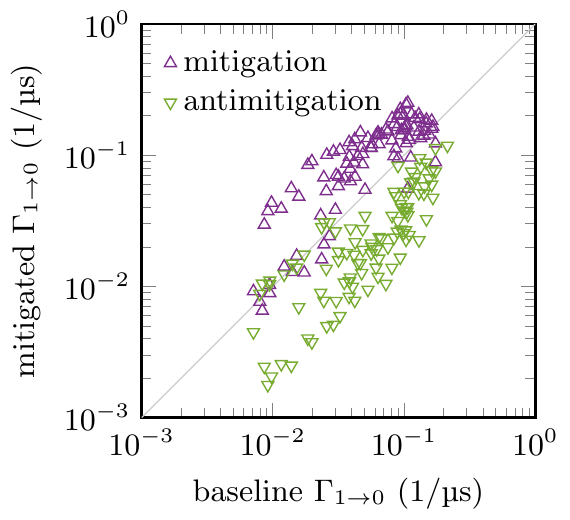}
\caption{\label{fig:escape-rates} Measurements of escape rates out of the first excited state. For each instance, we prepared the first excited state and measured the escape rate out of this state at the annealing parameter $s_{\rm target}$ that corresponded to the location of the anticrossing. We show data for baseline $\alpha=0$ and mitigated $\alpha = \pm 0.02$.}
\end{figure}

\subsection{Metastable valley escape rates}

To further illustrate the effect of the mitigaiton algorithm, we study the dynamics of relaxation from an excited state during the QA process with and without mitigation.  We prepare the QA hardware in a particular classical state at $s=1$ and run the QA algorithm {\em backwards} to an intermediate $s_{\rm target}$, where we remain for some dwell time $\tau$ before completing the QA algorithm back to $s = 1$. For a given classical state preparation, we can measure the initial escape rate $\Gamma$ out of this state at $s_{\rm target}$.

This measurement protocol is a generalization of the macroscopic resonant tunneling (MRT) protocol described elsewhere~\cite{Harris2008}. We measure $\Gamma$ for excited-state preparations for the uniform QA Hamiltonian and for the mitigated and antimitigated case. For each measurement, we choose an $s_{\rm target}$ that corresponds to the anticrossing identified by exact diagonalization of Hamitonian~\ref{eq:generalized}. Fig.~\ref{fig:escape-rates} shows the rate of escape from the metastable valley for the 100 instances.  Mitigation systematically enhances this rate and antimitigation suppresses it.

\section{Sampling degenerate ground states}

The initial testbed of 100 instances was designed to exhibit perturbative anticrossings late in the anneal.  Now we consider a second testbed of 100 24-qubit instances with highly degenerate ground states in which some ground states are observed much more frequently than others. In these instances deflection of eigenvalues late in the anneal leads to nonuniform sampling of classical states by QA \cite{Matsuda2009,Mandra2017,Zhang2017a}---this deflection does not lead to perturbative anticrossing because the eigenvalues converge at $s=1$.  Again the same mitigation approach---which does not consider whether or not observed states are excited---gives both empirical and spectral improvement.

\begin{figure}
  \setlength{\figurewidth}{2.8cm}
  \setlength{\figureheight}{2.8cm}
\includegraphics{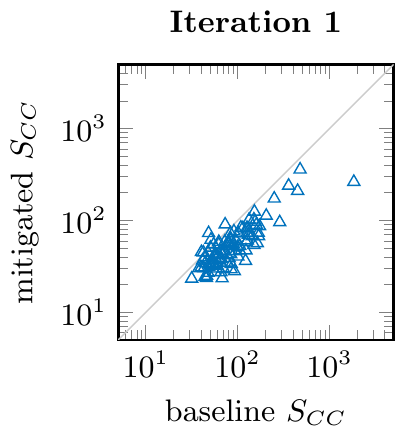}\includegraphics{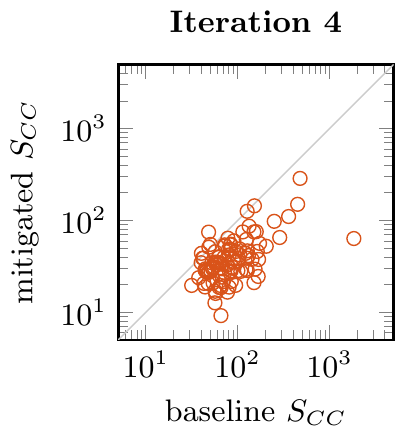}
\caption{\label{fig:sampling} Algorithmic mitigation improves uniformity of QA ground state sampling.  Shown are the expected number of samples required to observe every ground state, divided by the expected number for equal sampling so that each instance has minimum value 1.  Expectation values for 100 instances are shown after zero, one, and four iterations of mitigation.}
\end{figure}

With the optimization testbed, our performance metric was ground state probability $p_\mathrm{GS}$.  Here, with the {\em sampling testbed}, our performance metric is the expected number of samples required to see every ground state (up to antipodal symmetry) given by the empirical probabilities of each ground state, and normalized on a per-instance basis by the same expected value given perfectly uniform sampling of all ground states.  This metric is based on the {\em coupon collector} problem, and we denote it $S_{\mathrm{CC}}$ \cite{Flajolet1992, fclpaper}.  Fig.~\ref{fig:sampling} shows the results analogous to Fig.~\ref{fig:pgs-results}.  As before we see a systematic improvement in performance.

\begin{figure}
  \setlength{\figurewidth}{4cm}
  \setlength{\figureheight}{4cm}
  \includegraphics{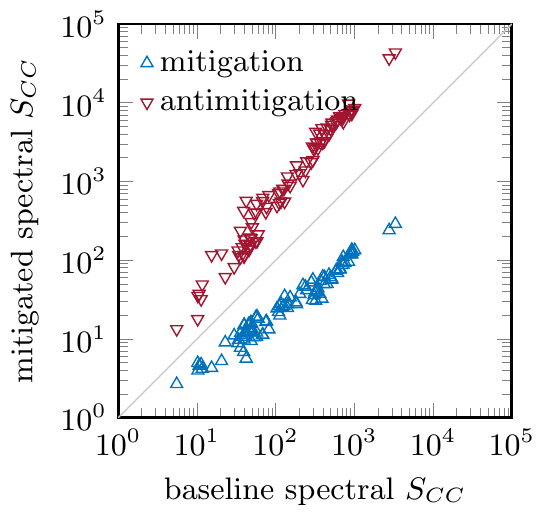}
  \caption{\label{fig:sampling_spectral}Algorithmic mitigation systematically improves uniformity of zero-temperature quantum Boltzmann distribution over classical ground states at $s=0.30$.}
\end{figure}

To measure spectral bias among classical ground states, we look at the support of the mid-anneal instantaneous ground state in the computational basis, specifically those basis vectors corresponding to classical ground states.  Where from an empirical perspective we can view classical ground state probabilities output by QA, from a spectral perspective we can view probabilities of classical ground states in the zero-temperature quantum Boltzmann distribution at some fixed point $s$ in the anneal.  In Fig.~\ref{fig:sampling_spectral} we show this data for $s=0.3$, contrasting the choice of $s=1$ in Ref.~\cite{Zhang2017a}; if we increase $s$ to $0.4$ the results look qualitatively similar but with larger $S_\mathrm{CC}$.

\section{Outstanding questions}

This work demonstrates a simple approach to a computational bottleneck in quantum annealing.  To move towards a more general demonstration there are some outstanding questions that need to be addressed:

\begin{itemize}
  \item {\bf Orthogonal control:} Currently available QA systems do not allow independent adjustment of $A(s)$ and $B(s)$; there is distortion of the classical Hamiltonian when $\delta_i \neq 0$.  This distortion may be helpful or harmful; we attempt to compensate for it in Section \ref{sec:orthogonal} and show that our approximation to transverse-only mitigation is successful for the instances studied.  The success of this initial demonstration highlights the need for future QA hardware to allow for at least partially independent adjustment of $A(s)$ and $B(s)$.

  \item {\bf Higher-order expansion:} We have used a first-order perturbative expansion to estimate the contribution of a small transverse field; this estimate is zero in the absence of floppy qubits.  In other problem classes such as Maximum Stable Set \cite{Dickson2012}, disordered Ising spin glasses \cite{Katzgraber2015}, or binary Ising spin glasses on graphs with odd connectivity \cite{Matsuda2009,tailspaper}, higher-order perturbative expansion is required because the first order term is zero.  The corresponding second-order algorithm has been shown to work in simulation \cite{Dickson2012} but remains to be studied in greater detail.

  \item {\bf Multiple avoided crossings:} The instances studied in this work have a relatively simple spectrum with a single avoided crossing. The mitigation algorithm effectively lifts this avoided crossing. In more complicated instances there may be multiple avoided crossings at different points during the QA algorithm~\cite{Knysh2016}.  A single iteration of the mitigation algorithm may alleviate multiple avoided crossings but if the number of avoided crossings increases quickly with problem size or first-order perturbation is insufficient to resolve all (or any) bottlenecks, the algorithm may fail to boost ground state probability.

  \item {\bf Offset granularity in large systems:} For this work we focused on small instances with clear avoided crossings and eigenspectra that are tractable to calculate throughout the QA algorithm.  At this problem scale the granularity of available anneal offsets---$\delta_i$ values are quantized at steps of approximately $0.002$---is not a barrier to success.  Larger systems, in addition to facing the attendant issues mentioned already, may require finer control of anneal offsets.

  \item {\bf Computational advantage:} This mitigation technique must ultimately confer a computational advantage in an algorithmic application in order to motivate practical use.  Our choice of 24-qubit instances allowed us to study the eigenspectra of the input instances, but strong evidence for a computational advantage would require the study of larger and more difficult inputs.

\end{itemize}

\section{Conclusions}

We have experimentally demonstrated an algorithm for mitigating perturbative anticrossings in quantum annealing.  The algorithm uses anneal offsets to slow the dynamics of floppy qubits, reducing the associated splitting of energy levels that causes perturbative anticrossings.  This results in an improvement of optimization performance in 100 small spin-glass instances designed to exhibit a response to this first-order perturbative mitigation.  Exact calculations of eigenspectra throughout the anneal confirm that this algorithm increases the eigengap and antimitigating reduces the eigengap.  Dynamics measurements confirm that under mitigation, relaxation out of excited states is enhanced; under antimitigation the opposite holds.

This work is the first experimental demonstration that a key computational bottleneck for quantum annealing---perturbative anticrossings---can be removed or mitigated with a targeted change in driving transverse Hamiltonian.  Extending these ideas to more complex input instances and more sophisticated mitigation algorithms will be the subject of future research.

\section{Acknowledgments}

We thank Richard Harris, Evgeny Andriyash, Mohammad Amin, Jack Raymond, and others at D-Wave for helpful conversations on this work and related topics.

\bibliography{bibtex}

\begin{thebibliography}{26}%
\makeatletter
\providecommand \@ifxundefined [1]{%
 \@ifx{#1\undefined}
}%
\providecommand \@ifnum [1]{%
 \ifnum #1\expandafter \@firstoftwo
 \else \expandafter \@secondoftwo
 \fi
}%
\providecommand \@ifx [1]{%
 \ifx #1\expandafter \@firstoftwo
 \else \expandafter \@secondoftwo
 \fi
}%
\providecommand \natexlab [1]{#1}%
\providecommand \enquote  [1]{``#1''}%
\providecommand \bibnamefont  [1]{#1}%
\providecommand \bibfnamefont [1]{#1}%
\providecommand \citenamefont [1]{#1}%
\providecommand \href@noop [0]{\@secondoftwo}%
\providecommand \href [0]{\begingroup \@sanitize@url \@href}%
\providecommand \@href[1]{\@@startlink{#1}\@@href}%
\providecommand \@@href[1]{\endgroup#1\@@endlink}%
\providecommand \@sanitize@url [0]{\catcode `\\12\catcode `\$12\catcode
  `\&12\catcode `\#12\catcode `\^12\catcode `\_12\catcode `\%12\relax}%
\providecommand \@@startlink[1]{}%
\providecommand \@@endlink[0]{}%
\providecommand \url  [0]{\begingroup\@sanitize@url \@url }%
\providecommand \@url [1]{\endgroup\@href {#1}{\urlprefix }}%
\providecommand \urlprefix  [0]{URL }%
\providecommand \Eprint [0]{\href }%
\providecommand \doibase [0]{http://dx.doi.org/}%
\providecommand \selectlanguage [0]{\@gobble}%
\providecommand \bibinfo  [0]{\@secondoftwo}%
\providecommand \bibfield  [0]{\@secondoftwo}%
\providecommand \translation [1]{[#1]}%
\providecommand \BibitemOpen [0]{}%
\providecommand \bibitemStop [0]{}%
\providecommand \bibitemNoStop [0]{.\EOS\space}%
\providecommand \EOS [0]{\spacefactor3000\relax}%
\providecommand \BibitemShut  [1]{\csname bibitem#1\endcsname}%
\let\auto@bib@innerbib\@empty
\bibitem [{\citenamefont {Lanting}\ \emph {et~al.}(2014)\citenamefont
  {Lanting}, \citenamefont {Przybysz}, \citenamefont {Smirnov}, \citenamefont
  {Spedalieri}, \citenamefont {Amin}, \citenamefont {Berkley}, \citenamefont
  {Harris}, \citenamefont {Altomare}, \citenamefont {Boixo}, \citenamefont
  {Bunyk}, \citenamefont {Dickson}, \citenamefont {Enderud}, \citenamefont
  {Hilton}, \citenamefont {Hoskinson}, \citenamefont {Johnson}, \citenamefont
  {Ladizinsky}, \citenamefont {Ladizinsky}, \citenamefont {Neufeld},
  \citenamefont {Oh}, \citenamefont {Perminov}, \citenamefont {Rich},
  \citenamefont {Thom}, \citenamefont {Tolkacheva}, \citenamefont {Uchaikin},
  \citenamefont {Wilson},\ and\ \citenamefont {Rose}}]{Lanting2014}%
  \BibitemOpen
  \bibfield  {author} {\bibinfo {author} {\bibfnamefont {T.}~\bibnamefont
  {Lanting}}, \bibinfo {author} {\bibfnamefont {A.~J.}\ \bibnamefont
  {Przybysz}}, \bibinfo {author} {\bibfnamefont {A.~Y.}\ \bibnamefont
  {Smirnov}}, \bibinfo {author} {\bibfnamefont {F.~M.}\ \bibnamefont
  {Spedalieri}}, \bibinfo {author} {\bibfnamefont {M.~H.}\ \bibnamefont
  {Amin}}, \bibinfo {author} {\bibfnamefont {A.~J.}\ \bibnamefont {Berkley}},
  \bibinfo {author} {\bibfnamefont {R.}~\bibnamefont {Harris}}, \bibinfo
  {author} {\bibfnamefont {F.}~\bibnamefont {Altomare}}, \bibinfo {author}
  {\bibfnamefont {S.}~\bibnamefont {Boixo}}, \bibinfo {author} {\bibfnamefont
  {P.}~\bibnamefont {Bunyk}}, \bibinfo {author} {\bibfnamefont
  {N.}~\bibnamefont {Dickson}}, \bibinfo {author} {\bibfnamefont
  {C.}~\bibnamefont {Enderud}}, \bibinfo {author} {\bibfnamefont {J.~P.}\
  \bibnamefont {Hilton}}, \bibinfo {author} {\bibfnamefont {E.}~\bibnamefont
  {Hoskinson}}, \bibinfo {author} {\bibfnamefont {M.~W.}\ \bibnamefont
  {Johnson}}, \bibinfo {author} {\bibfnamefont {E.}~\bibnamefont {Ladizinsky}},
  \bibinfo {author} {\bibfnamefont {N.}~\bibnamefont {Ladizinsky}}, \bibinfo
  {author} {\bibfnamefont {R.}~\bibnamefont {Neufeld}}, \bibinfo {author}
  {\bibfnamefont {T.}~\bibnamefont {Oh}}, \bibinfo {author} {\bibfnamefont
  {I.}~\bibnamefont {Perminov}}, \bibinfo {author} {\bibfnamefont
  {C.}~\bibnamefont {Rich}}, \bibinfo {author} {\bibfnamefont {M.~C.}\
  \bibnamefont {Thom}}, \bibinfo {author} {\bibfnamefont {E.}~\bibnamefont
  {Tolkacheva}}, \bibinfo {author} {\bibfnamefont {S.}~\bibnamefont
  {Uchaikin}}, \bibinfo {author} {\bibfnamefont {A.~B.}\ \bibnamefont
  {Wilson}}, \ and\ \bibinfo {author} {\bibfnamefont {G.}~\bibnamefont
  {Rose}},\ }\href {\doibase 10.1103/PhysRevX.4.021041} {\bibfield  {journal}
  {\bibinfo  {journal} {Physical Review X}\ }\textbf {\bibinfo {volume} {4}},\
  \bibinfo {pages} {021041} (\bibinfo {year} {2014})}\BibitemShut {NoStop}%
\bibitem [{\citenamefont {Denchev}\ \emph {et~al.}(2016)\citenamefont
  {Denchev}, \citenamefont {Boixo}, \citenamefont {Isakov}, \citenamefont
  {Ding}, \citenamefont {Babbush}, \citenamefont {Smelyanskiy}, \citenamefont
  {Martinis},\ and\ \citenamefont {Neven}}]{Denchev2016}%
  \BibitemOpen
  \bibfield  {author} {\bibinfo {author} {\bibfnamefont {V.~S.}\ \bibnamefont
  {Denchev}}, \bibinfo {author} {\bibfnamefont {S.}~\bibnamefont {Boixo}},
  \bibinfo {author} {\bibfnamefont {S.~V.}\ \bibnamefont {Isakov}}, \bibinfo
  {author} {\bibfnamefont {N.}~\bibnamefont {Ding}}, \bibinfo {author}
  {\bibfnamefont {R.}~\bibnamefont {Babbush}}, \bibinfo {author} {\bibfnamefont
  {V.}~\bibnamefont {Smelyanskiy}}, \bibinfo {author} {\bibfnamefont
  {J.}~\bibnamefont {Martinis}}, \ and\ \bibinfo {author} {\bibfnamefont
  {H.}~\bibnamefont {Neven}},\ }\href {\doibase 10.1103/PhysRevX.6.031015}
  {\bibfield  {journal} {\bibinfo  {journal} {Phys. Rev. X}\ }\textbf {\bibinfo
  {volume} {6}},\ \bibinfo {pages} {031015} (\bibinfo {year}
  {2016})}\BibitemShut {NoStop}%
\bibitem [{\citenamefont {Andriyash}\ and\ \citenamefont
  {Amin}(2017)}]{Andriyash2017}%
  \BibitemOpen
  \bibfield  {author} {\bibinfo {author} {\bibfnamefont {E.}~\bibnamefont
  {Andriyash}}\ and\ \bibinfo {author} {\bibfnamefont {M.~H.}\ \bibnamefont
  {Amin}},\ }\href {http://arxiv.org/abs/1703.09277} {\enquote {\bibinfo
  {title} {{Can quantum Monte Carlo simulate quantum annealing?}}}\ } (\bibinfo
  {year} {2017}),\ \Eprint {http://arxiv.org/abs/1703.09277} {arXiv:1703.09277}
  \BibitemShut {NoStop}%
\bibitem [{\citenamefont {Boixo}\ \emph {et~al.}(2013)\citenamefont {Boixo},
  \citenamefont {Albash}, \citenamefont {Spedalieri}, \citenamefont
  {Chancellor},\ and\ \citenamefont {Lidar}}]{Boixo2013}%
  \BibitemOpen
  \bibfield  {author} {\bibinfo {author} {\bibfnamefont {S.}~\bibnamefont
  {Boixo}}, \bibinfo {author} {\bibfnamefont {T.}~\bibnamefont {Albash}},
  \bibinfo {author} {\bibfnamefont {F.~M.}\ \bibnamefont {Spedalieri}},
  \bibinfo {author} {\bibfnamefont {N.}~\bibnamefont {Chancellor}}, \ and\
  \bibinfo {author} {\bibfnamefont {D.~a.}\ \bibnamefont {Lidar}},\ }\href
  {\doibase 10.1038/ncomms3067} {\bibfield  {journal} {\bibinfo  {journal}
  {Nature Communications}\ }\textbf {\bibinfo {volume} {4}},\ \bibinfo {pages}
  {2067} (\bibinfo {year} {2013})}\BibitemShut {NoStop}%
\bibitem [{\citenamefont {Steiger}\ \emph {et~al.}(2015)\citenamefont
  {Steiger}, \citenamefont {R\o{}nnow},\ and\ \citenamefont
  {Troyer}}]{Steiger2015}%
  \BibitemOpen
  \bibfield  {author} {\bibinfo {author} {\bibfnamefont {D.~S.}\ \bibnamefont
  {Steiger}}, \bibinfo {author} {\bibfnamefont {T.~F.}\ \bibnamefont
  {R\o{}nnow}}, \ and\ \bibinfo {author} {\bibfnamefont {M.}~\bibnamefont
  {Troyer}},\ }\href {\doibase 10.1103/PhysRevLett.115.230501} {\bibfield
  {journal} {\bibinfo  {journal} {Phys. Rev. Lett.}\ }\textbf {\bibinfo
  {volume} {115}},\ \bibinfo {pages} {230501} (\bibinfo {year}
  {2015})}\BibitemShut {NoStop}%
\bibitem [{\citenamefont {King}\ \emph {et~al.}(2016)\citenamefont {King},
  \citenamefont {Hoskinson}, \citenamefont {Lanting}, \citenamefont
  {Andriyash},\ and\ \citenamefont {Amin}}]{tailspaper}%
  \BibitemOpen
  \bibfield  {author} {\bibinfo {author} {\bibfnamefont {A.~D.}\ \bibnamefont
  {King}}, \bibinfo {author} {\bibfnamefont {E.}~\bibnamefont {Hoskinson}},
  \bibinfo {author} {\bibfnamefont {T.}~\bibnamefont {Lanting}}, \bibinfo
  {author} {\bibfnamefont {E.}~\bibnamefont {Andriyash}}, \ and\ \bibinfo
  {author} {\bibfnamefont {M.~H.}\ \bibnamefont {Amin}},\ }\href {\doibase
  10.1103/PhysRevA.93.052320} {\bibfield  {journal} {\bibinfo  {journal}
  {Physical Review A}\ }\textbf {\bibinfo {volume} {93}},\ \bibinfo {pages}
  {052320} (\bibinfo {year} {2016})}\BibitemShut {NoStop}%
\bibitem [{\citenamefont {Dickson}\ and\ \citenamefont
  {Amin}(2012)}]{Dickson2012}%
  \BibitemOpen
  \bibfield  {author} {\bibinfo {author} {\bibfnamefont {N.~G.}\ \bibnamefont
  {Dickson}}\ and\ \bibinfo {author} {\bibfnamefont {M.~H.}\ \bibnamefont
  {Amin}},\ }\href {\doibase 10.1103/PhysRevA.85.032303} {\bibfield  {journal}
  {\bibinfo  {journal} {Physical Review A}\ }\textbf {\bibinfo {volume} {85}},\
  \bibinfo {pages} {032303} (\bibinfo {year} {2012})}\BibitemShut {NoStop}%
\bibitem [{\citenamefont {Boixo}\ \emph {et~al.}(2014)\citenamefont {Boixo},
  \citenamefont {R{\o}nnow}, \citenamefont {Isakov}, \citenamefont {Wang},
  \citenamefont {Wecker}, \citenamefont {Lidar}, \citenamefont {Martinis},\
  and\ \citenamefont {Troyer}}]{boixo2014evidence}%
  \BibitemOpen
  \bibfield  {author} {\bibinfo {author} {\bibfnamefont {S.}~\bibnamefont
  {Boixo}}, \bibinfo {author} {\bibfnamefont {T.~F.}\ \bibnamefont
  {R{\o}nnow}}, \bibinfo {author} {\bibfnamefont {S.~V.}\ \bibnamefont
  {Isakov}}, \bibinfo {author} {\bibfnamefont {Z.}~\bibnamefont {Wang}},
  \bibinfo {author} {\bibfnamefont {D.}~\bibnamefont {Wecker}}, \bibinfo
  {author} {\bibfnamefont {D.~A.}\ \bibnamefont {Lidar}}, \bibinfo {author}
  {\bibfnamefont {J.~M.}\ \bibnamefont {Martinis}}, \ and\ \bibinfo {author}
  {\bibfnamefont {M.}~\bibnamefont {Troyer}},\ }\href {\doibase
  10.1038/nphys2900} {\bibfield  {journal} {\bibinfo  {journal} {Nature
  Physics}\ }\textbf {\bibinfo {volume} {10}},\ \bibinfo {pages} {218}
  (\bibinfo {year} {2014})}\BibitemShut {NoStop}%
\bibitem [{\citenamefont {{Amin}}\ and\ \citenamefont
  {{Choi}}(2009)}]{amin-choi}%
  \BibitemOpen
  \bibfield  {author} {\bibinfo {author} {\bibfnamefont {M.~H.~S.}\
  \bibnamefont {{Amin}}}\ and\ \bibinfo {author} {\bibfnamefont
  {V.}~\bibnamefont {{Choi}}},\ }\href {\doibase 10.1103/PhysRevA.80.062326}
  {\bibfield  {journal} {\bibinfo  {journal} {\pra}\ }\textbf {\bibinfo
  {volume} {80}},\ \bibinfo {eid} {062326} (\bibinfo {year} {2009})},\ \Eprint
  {http://arxiv.org/abs/0904.1387} {arXiv:0904.1387 [quant-ph]} \BibitemShut
  {NoStop}%
\bibitem [{\citenamefont {Dickson}\ \emph {et~al.}(2013)\citenamefont {Dickson}
  \emph {et~al.}}]{Dickson2013}%
  \BibitemOpen
  \bibfield  {author} {\bibinfo {author} {\bibfnamefont {N.}~\bibnamefont
  {Dickson}} \emph {et~al.},\ }\href {\doibase 10.1038/ncomms2920} {\bibfield
  {journal} {\bibinfo  {journal} {Nat. Commun.}\ }\textbf {\bibinfo {volume}
  {4}},\ \bibinfo {pages} {1903} (\bibinfo {year} {2013})}\BibitemShut
  {NoStop}%
\bibitem [{\citenamefont {Albash}\ \emph {et~al.}(2015)\citenamefont {Albash},
  \citenamefont {Vinci}, \citenamefont {Mishra}, \citenamefont {Warburton},\
  and\ \citenamefont {Lidar}}]{Albash2015}%
  \BibitemOpen
  \bibfield  {author} {\bibinfo {author} {\bibfnamefont {T.}~\bibnamefont
  {Albash}}, \bibinfo {author} {\bibfnamefont {W.}~\bibnamefont {Vinci}},
  \bibinfo {author} {\bibfnamefont {A.}~\bibnamefont {Mishra}}, \bibinfo
  {author} {\bibfnamefont {P.~a.}\ \bibnamefont {Warburton}}, \ and\ \bibinfo
  {author} {\bibfnamefont {D.~A.}\ \bibnamefont {Lidar}},\ }\href {\doibase
  10.1103/PhysRevA.91.042314} {\bibfield  {journal} {\bibinfo  {journal}
  {Physical Review A}\ }\textbf {\bibinfo {volume} {91}},\ \bibinfo {pages}
  {042314} (\bibinfo {year} {2015})}\BibitemShut {NoStop}%
\bibitem [{\citenamefont {Harris}\ \emph {et~al.}(2008)\citenamefont {Harris},
  \citenamefont {Johnson}, \citenamefont {Han}, \citenamefont {Berkley},
  \citenamefont {Johansson}, \citenamefont {Bunyk}, \citenamefont {Ladizinsky},
  \citenamefont {Govorkov}, \citenamefont {Thom}, \citenamefont {Uchaikin},
  \citenamefont {Bumble}, \citenamefont {Fung}, \citenamefont {Kaul},
  \citenamefont {Kleinsasser}, \citenamefont {Amin},\ and\ \citenamefont
  {Averin}}]{Harris2008}%
  \BibitemOpen
  \bibfield  {author} {\bibinfo {author} {\bibfnamefont {R.}~\bibnamefont
  {Harris}}, \bibinfo {author} {\bibfnamefont {M.~W.}\ \bibnamefont {Johnson}},
  \bibinfo {author} {\bibfnamefont {S.}~\bibnamefont {Han}}, \bibinfo {author}
  {\bibfnamefont {A.~J.}\ \bibnamefont {Berkley}}, \bibinfo {author}
  {\bibfnamefont {J.}~\bibnamefont {Johansson}}, \bibinfo {author}
  {\bibfnamefont {P.}~\bibnamefont {Bunyk}}, \bibinfo {author} {\bibfnamefont
  {E.}~\bibnamefont {Ladizinsky}}, \bibinfo {author} {\bibfnamefont
  {S.}~\bibnamefont {Govorkov}}, \bibinfo {author} {\bibfnamefont {M.~C.}\
  \bibnamefont {Thom}}, \bibinfo {author} {\bibfnamefont {S.}~\bibnamefont
  {Uchaikin}}, \bibinfo {author} {\bibfnamefont {B.}~\bibnamefont {Bumble}},
  \bibinfo {author} {\bibfnamefont {A.}~\bibnamefont {Fung}}, \bibinfo {author}
  {\bibfnamefont {A.}~\bibnamefont {Kaul}}, \bibinfo {author} {\bibfnamefont
  {A.}~\bibnamefont {Kleinsasser}}, \bibinfo {author} {\bibfnamefont
  {M.~H.~S.}\ \bibnamefont {Amin}}, \ and\ \bibinfo {author} {\bibfnamefont
  {D.~V.}\ \bibnamefont {Averin}},\ }\href {\doibase
  10.1103/PhysRevLett.101.117003} {\bibfield  {journal} {\bibinfo  {journal}
  {Physical Review Letters}\ }\textbf {\bibinfo {volume} {101}},\ \bibinfo
  {pages} {117003} (\bibinfo {year} {2008})}\BibitemShut {NoStop}%
\bibitem [{\citenamefont {Matsuda}\ \emph {et~al.}(2009)\citenamefont
  {Matsuda}, \citenamefont {Nishimori},\ and\ \citenamefont
  {Katzgraber}}]{Matsuda2009}%
  \BibitemOpen
  \bibfield  {author} {\bibinfo {author} {\bibfnamefont {Y.}~\bibnamefont
  {Matsuda}}, \bibinfo {author} {\bibfnamefont {H.}~\bibnamefont {Nishimori}},
  \ and\ \bibinfo {author} {\bibfnamefont {H.~G.}\ \bibnamefont {Katzgraber}},\
  }\href {\doibase 10.1088/1367-2630/11/7/073021} {\bibfield  {journal}
  {\bibinfo  {journal} {New Journal of Physics}\ }\textbf {\bibinfo {volume}
  {11}},\ \bibinfo {pages} {073021} (\bibinfo {year} {2009})}\BibitemShut
  {NoStop}%
\bibitem [{\citenamefont {Mandr{\`{a}}}\ \emph {et~al.}(2017)\citenamefont
  {Mandr{\`{a}}}, \citenamefont {Zhu},\ and\ \citenamefont
  {Katzgraber}}]{Mandra2017}%
  \BibitemOpen
  \bibfield  {author} {\bibinfo {author} {\bibfnamefont {S.}~\bibnamefont
  {Mandr{\`{a}}}}, \bibinfo {author} {\bibfnamefont {Z.}~\bibnamefont {Zhu}}, \
  and\ \bibinfo {author} {\bibfnamefont {H.~G.}\ \bibnamefont {Katzgraber}},\
  }\href {\doibase 10.1103/PhysRevLett.118.070502} {\bibfield  {journal}
  {\bibinfo  {journal} {Physical Review Letters}\ }\textbf {\bibinfo {volume}
  {118}},\ \bibinfo {pages} {070502} (\bibinfo {year} {2017})}\BibitemShut
  {NoStop}%
\bibitem [{\citenamefont {Zhang}\ \emph {et~al.}(2017)\citenamefont {Zhang},
  \citenamefont {Wagenbreth}, \citenamefont {Martin-Mayor},\ and\ \citenamefont
  {Hen}}]{Zhang2017a}%
  \BibitemOpen
  \bibfield  {author} {\bibinfo {author} {\bibfnamefont {B.~H.}\ \bibnamefont
  {Zhang}}, \bibinfo {author} {\bibfnamefont {G.}~\bibnamefont {Wagenbreth}},
  \bibinfo {author} {\bibfnamefont {V.}~\bibnamefont {Martin-Mayor}}, \ and\
  \bibinfo {author} {\bibfnamefont {I.}~\bibnamefont {Hen}},\ }\href {\doibase
  10.1038/s41598-017-01096-6} {\bibfield  {journal} {\bibinfo  {journal}
  {Scientific Reports}\ }\textbf {\bibinfo {volume} {7}},\ \bibinfo {pages}
  {1044} (\bibinfo {year} {2017})}\BibitemShut {NoStop}%
\bibitem [{\citenamefont {Flajolet}\ \emph {et~al.}(1992)\citenamefont
  {Flajolet}, \citenamefont {Gardy},\ and\ \citenamefont
  {Thimonier}}]{Flajolet1992}%
  \BibitemOpen
  \bibfield  {author} {\bibinfo {author} {\bibfnamefont {P.}~\bibnamefont
  {Flajolet}}, \bibinfo {author} {\bibfnamefont {D.}~\bibnamefont {Gardy}}, \
  and\ \bibinfo {author} {\bibfnamefont {L.}~\bibnamefont {Thimonier}},\ }\href
  {\doibase 10.1016/0166-218X(92)90177-C} {\bibfield  {journal} {\bibinfo
  {journal} {Discrete Applied Mathematics}\ }\textbf {\bibinfo {volume} {39}},\
  \bibinfo {pages} {207} (\bibinfo {year} {1992})}\BibitemShut {NoStop}%
\bibitem [{\citenamefont {King}\ \emph {et~al.}(2017)\citenamefont {King},
  \citenamefont {Yarkoni}, \citenamefont {Raymond}, \citenamefont {Ozfidan},
  \citenamefont {King}, \citenamefont {Nevisi}, \citenamefont {Hilton},\ and\
  \citenamefont {McGeoch}}]{fclpaper}%
  \BibitemOpen
  \bibfield  {author} {\bibinfo {author} {\bibfnamefont {J.}~\bibnamefont
  {King}}, \bibinfo {author} {\bibfnamefont {S.}~\bibnamefont {Yarkoni}},
  \bibinfo {author} {\bibfnamefont {J.}~\bibnamefont {Raymond}}, \bibinfo
  {author} {\bibfnamefont {I.}~\bibnamefont {Ozfidan}}, \bibinfo {author}
  {\bibfnamefont {A.~D.}\ \bibnamefont {King}}, \bibinfo {author}
  {\bibfnamefont {M.~M.}\ \bibnamefont {Nevisi}}, \bibinfo {author}
  {\bibfnamefont {J.~P.}\ \bibnamefont {Hilton}}, \ and\ \bibinfo {author}
  {\bibfnamefont {C.~C.}\ \bibnamefont {McGeoch}},\ }\href
  {http://arxiv.org/abs/1701.04579} {\enquote {\bibinfo {title} {{Quantum
  Annealing amid Local Ruggedness and Global Frustration}},}\ } (\bibinfo
  {year} {2017}),\ \Eprint {http://arxiv.org/abs/1701.04579} {arXiv:1701.04579}
  \BibitemShut {NoStop}%
\bibitem [{\citenamefont {Katzgraber}\ \emph {et~al.}(2015)\citenamefont
  {Katzgraber}, \citenamefont {Hamze}, \citenamefont {Zhu}, \citenamefont
  {Ochoa},\ and\ \citenamefont {Munoz-Bauza}}]{Katzgraber2015}%
  \BibitemOpen
  \bibfield  {author} {\bibinfo {author} {\bibfnamefont {H.~G.}\ \bibnamefont
  {Katzgraber}}, \bibinfo {author} {\bibfnamefont {F.}~\bibnamefont {Hamze}},
  \bibinfo {author} {\bibfnamefont {Z.}~\bibnamefont {Zhu}}, \bibinfo {author}
  {\bibfnamefont {A.~J.}\ \bibnamefont {Ochoa}}, \ and\ \bibinfo {author}
  {\bibfnamefont {H.}~\bibnamefont {Munoz-Bauza}},\ }\href {\doibase
  10.1103/PhysRevX.5.031026} {\bibfield  {journal} {\bibinfo  {journal} {Phys.
  Rev. X}\ }\textbf {\bibinfo {volume} {5}},\ \bibinfo {pages} {031026}
  (\bibinfo {year} {2015})}\BibitemShut {NoStop}%
\bibitem [{\citenamefont {Knysh}(2016)}]{Knysh2016}%
  \BibitemOpen
  \bibfield  {author} {\bibinfo {author} {\bibfnamefont {S.}~\bibnamefont
  {Knysh}},\ }\href {\doibase 10.1038/ncomms12370} {\bibfield  {journal}
  {\bibinfo  {journal} {Nature Communications}\ }\textbf {\bibinfo {volume}
  {7}},\ \bibinfo {pages} {12370} (\bibinfo {year} {2016})}\BibitemShut
  {NoStop}%
\bibitem [{\citenamefont {Harris}\ \emph {et~al.}(2010)\citenamefont {Harris},
  \citenamefont {Johansson}, \citenamefont {Berkley}, \citenamefont {Johnson},
  \citenamefont {Lanting}, \citenamefont {Han}, \citenamefont {Bunyk},
  \citenamefont {Ladizinsky}, \citenamefont {Oh}, \citenamefont {Perminov},
  \citenamefont {Tolkacheva}, \citenamefont {Uchaikin}, \citenamefont
  {Chapple}, \citenamefont {Enderud}, \citenamefont {Rich}, \citenamefont
  {Thom}, \citenamefont {Wang}, \citenamefont {Wilson},\ and\ \citenamefont
  {Rose}}]{Harris2010}%
  \BibitemOpen
  \bibfield  {author} {\bibinfo {author} {\bibfnamefont {R.}~\bibnamefont
  {Harris}}, \bibinfo {author} {\bibfnamefont {J.}~\bibnamefont {Johansson}},
  \bibinfo {author} {\bibfnamefont {A.~J.}\ \bibnamefont {Berkley}}, \bibinfo
  {author} {\bibfnamefont {M.~W.}\ \bibnamefont {Johnson}}, \bibinfo {author}
  {\bibfnamefont {T.}~\bibnamefont {Lanting}}, \bibinfo {author} {\bibfnamefont
  {S.}~\bibnamefont {Han}}, \bibinfo {author} {\bibfnamefont {P.}~\bibnamefont
  {Bunyk}}, \bibinfo {author} {\bibfnamefont {E.}~\bibnamefont {Ladizinsky}},
  \bibinfo {author} {\bibfnamefont {T.}~\bibnamefont {Oh}}, \bibinfo {author}
  {\bibfnamefont {I.}~\bibnamefont {Perminov}}, \bibinfo {author}
  {\bibfnamefont {E.}~\bibnamefont {Tolkacheva}}, \bibinfo {author}
  {\bibfnamefont {S.}~\bibnamefont {Uchaikin}}, \bibinfo {author}
  {\bibfnamefont {E.~M.}\ \bibnamefont {Chapple}}, \bibinfo {author}
  {\bibfnamefont {C.}~\bibnamefont {Enderud}}, \bibinfo {author} {\bibfnamefont
  {C.}~\bibnamefont {Rich}}, \bibinfo {author} {\bibfnamefont {M.}~\bibnamefont
  {Thom}}, \bibinfo {author} {\bibfnamefont {J.}~\bibnamefont {Wang}}, \bibinfo
  {author} {\bibfnamefont {B.}~\bibnamefont {Wilson}}, \ and\ \bibinfo {author}
  {\bibfnamefont {G.}~\bibnamefont {Rose}},\ }\href@noop {} {\bibfield
  {journal} {\bibinfo  {journal} {Phys. Rev. B}\ }\textbf {\bibinfo {volume}
  {81}},\ \bibinfo {pages} {134510} (\bibinfo {year} {2010})}\BibitemShut
  {NoStop}%
\bibitem [{\citenamefont {{Harris}}\ \emph {et~al.}(2009)\citenamefont
  {{Harris}}, \citenamefont {{Lanting}}, \citenamefont {{Berkley}},
  \citenamefont {{Johansson}}, \citenamefont {{Johnson}}, \citenamefont
  {{Bunyk}}, \citenamefont {{Ladizinsky}}, \citenamefont {{Ladizinsky}},
  \citenamefont {{Oh}},\ and\ \citenamefont {{Han}}}]{harris-cjc-2009}%
  \BibitemOpen
  \bibfield  {author} {\bibinfo {author} {\bibfnamefont {R.}~\bibnamefont
  {{Harris}}}, \bibinfo {author} {\bibfnamefont {T.}~\bibnamefont {{Lanting}}},
  \bibinfo {author} {\bibfnamefont {A.~J.}\ \bibnamefont {{Berkley}}}, \bibinfo
  {author} {\bibfnamefont {J.}~\bibnamefont {{Johansson}}}, \bibinfo {author}
  {\bibfnamefont {M.~W.}\ \bibnamefont {{Johnson}}}, \bibinfo {author}
  {\bibfnamefont {P.}~\bibnamefont {{Bunyk}}}, \bibinfo {author} {\bibfnamefont
  {E.}~\bibnamefont {{Ladizinsky}}}, \bibinfo {author} {\bibfnamefont
  {N.}~\bibnamefont {{Ladizinsky}}}, \bibinfo {author} {\bibfnamefont
  {T.}~\bibnamefont {{Oh}}}, \ and\ \bibinfo {author} {\bibfnamefont
  {S.}~\bibnamefont {{Han}}},\ }\href {\doibase 10.1103/PhysRevB.80.052506}
  {\bibfield  {journal} {\bibinfo  {journal} {Phys. Rev. B}\ }\textbf {\bibinfo
  {volume} {80}},\ \bibinfo {pages} {052506} (\bibinfo {year}
  {2009})}\BibitemShut {NoStop}%
\bibitem [{\citenamefont {King}\ and\ \citenamefont
  {McGeoch}(2014)}]{King2014}%
  \BibitemOpen
  \bibfield  {author} {\bibinfo {author} {\bibfnamefont {A.~D.}\ \bibnamefont
  {King}}\ and\ \bibinfo {author} {\bibfnamefont {C.~C.}\ \bibnamefont
  {McGeoch}},\ }\href@noop {} {\bibfield  {journal} {\bibinfo  {journal} {arXiv
  preprint arXiv:1410.2628}\ } (\bibinfo {year} {2014})}\BibitemShut {NoStop}%
\bibitem [{\citenamefont {Perron}(1907)}]{Perron1907}%
  \BibitemOpen
  \bibfield  {author} {\bibinfo {author} {\bibfnamefont {O.}~\bibnamefont
  {Perron}},\ }\href {http://eudml.org/doc/158317} {\bibfield  {journal}
  {\bibinfo  {journal} {Mathematische Annalen}\ }\textbf {\bibinfo {volume}
  {64}},\ \bibinfo {pages} {248} (\bibinfo {year} {1907})}\BibitemShut
  {NoStop}%
\bibitem [{\citenamefont {Farhi}\ \emph {et~al.}(2000)\citenamefont {Farhi},
  \citenamefont {Goldstone}, \citenamefont {Gutmann},\ and\ \citenamefont
  {Sipser}}]{Farhi2000}%
  \BibitemOpen
  \bibfield  {author} {\bibinfo {author} {\bibfnamefont {E.}~\bibnamefont
  {Farhi}}, \bibinfo {author} {\bibfnamefont {J.}~\bibnamefont {Goldstone}},
  \bibinfo {author} {\bibfnamefont {S.}~\bibnamefont {Gutmann}}, \ and\
  \bibinfo {author} {\bibfnamefont {M.}~\bibnamefont {Sipser}},\ }\href
  {http://arxiv.org/abs/quant-ph/0001106} {\enquote {\bibinfo {title} {{Quantum
  Computation by Adiabatic Evolution}},}\ } (\bibinfo {year} {2000}),\ \Eprint
  {http://arxiv.org/abs/0001106} {arXiv:0001106 [quant-ph]} \BibitemShut
  {NoStop}%
\bibitem [{\citenamefont {Farhi}\ \emph {et~al.}(2001)\citenamefont {Farhi},
  \citenamefont {Goldstone}, \citenamefont {Gutmann}, \citenamefont {Lapan},
  \citenamefont {Lundgren},\ and\ \citenamefont {Preda}}]{Farhi2001}%
  \BibitemOpen
  \bibfield  {author} {\bibinfo {author} {\bibfnamefont {E.}~\bibnamefont
  {Farhi}}, \bibinfo {author} {\bibfnamefont {J.}~\bibnamefont {Goldstone}},
  \bibinfo {author} {\bibfnamefont {S.}~\bibnamefont {Gutmann}}, \bibinfo
  {author} {\bibfnamefont {J.}~\bibnamefont {Lapan}}, \bibinfo {author}
  {\bibfnamefont {A.}~\bibnamefont {Lundgren}}, \ and\ \bibinfo {author}
  {\bibfnamefont {D.}~\bibnamefont {Preda}},\ }\href {\doibase
  10.1126/science.1057726} {\bibfield  {journal} {\bibinfo  {journal}
  {Science}\ }\textbf {\bibinfo {volume} {292}},\ \bibinfo {pages} {472}
  (\bibinfo {year} {2001})}\BibitemShut {NoStop}%
\bibitem [{\citenamefont {Smelyanskiy}\ \emph {et~al.}(2001)\citenamefont
  {Smelyanskiy}, \citenamefont {Toussaint},\ and\ \citenamefont
  {Timucin}}]{Smelyanskiy2001}%
  \BibitemOpen
  \bibfield  {author} {\bibinfo {author} {\bibfnamefont {V.~N.}\ \bibnamefont
  {Smelyanskiy}}, \bibinfo {author} {\bibfnamefont {U.~V.}\ \bibnamefont
  {Toussaint}}, \ and\ \bibinfo {author} {\bibfnamefont {D.~A.}\ \bibnamefont
  {Timucin}},\ }\href {http://arxiv.org/abs/quant-ph/0112143} {\enquote
  {\bibinfo {title} {{Simulations of the adiabatic quantum optimization for the
  Set Partition Problem}},}\ } (\bibinfo {year} {2001}),\ \Eprint
  {http://arxiv.org/abs/0112143} {arXiv:0112143 [quant-ph]} \BibitemShut
  {NoStop}%
\end{thebibliography}%

\appendix

\section{QA apparatus and methods}\label{app:methods}

We acquired the data reported in this manuscript with a D-Wave 2000Q system in Burnaby, British Columbia. Each qubit is a compound-compound Josephson-junction rf SQUID~\cite{Harris2010}. The total critical current of the rf-SQUID junctions is $2.56\ \mu$A, the rf-SQUID body inductance is $282$ pH, and the rf-SQUID capacitance is $98$ fF. The qubits are tunably coupled with inter-qubit magnetic coupling elements that produce a maximum anti-ferromagnetic, $M_{\rm AFM} = 2.12$ pH~\cite{harris-cjc-2009}.

We run the QA algorithm by adjusting the external bias on the compound-compound Josephson-junction loop $\Phi^x_{\rm CCJJ}$ from $\Phi^i_{\rm CCJJ}/\Phi_0 = -0.6457$ at $s = 0$ to $\Phi^f_{\rm CCJJ}/\Phi_0 = -0.7140$ at $s = 1$ according to the curve shown in Fig.~\ref{FIG:uniform-annealing}. We ran the QA algorithm with $t_f = \SI{20}{\micro\second}$ at a processor temperature of  $12.7\pm 0.5\SI{}{\milli\kelvin}$. Of the $2048$ qubits physically present, $2033$ were operational. This allowed us to embed $63$ disjoint copies of a $24$-qubit system on the processor, each subject to an independent spin reversal transformation \cite{King2014}.

\begin{figure}
\includegraphics[width=0.5\textwidth]{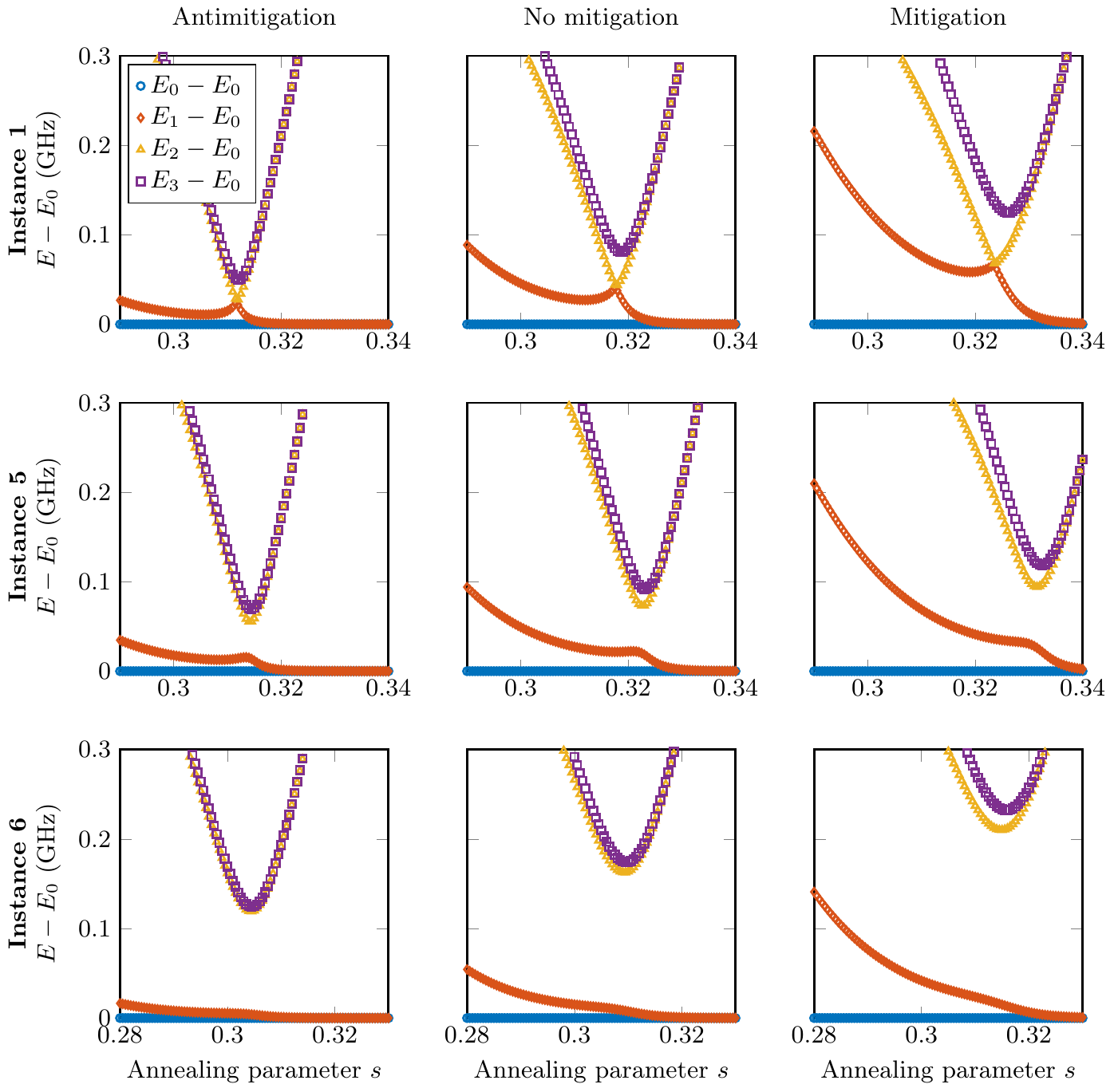}
  \caption{\label{fig:exemplars}First, second, and third eigengaps of three exemplary instances using antimitigation (left), no mitigation (middle), and mitigation (right).  First eigengap $E_1-E_0$ always goes to zero late in the anneal; the third eigengap gives a clean representation of the desired tunneling event between superpositions of classical ground and excited states.}
\end{figure}

\section{Perturbation calculation details}\label{app:perturbation}

Consider a system with $N$ degenerate eigenstates with energy $E_1$ connected to one another via single bit flips. The presence of a transverse field $A(s)$ lifts this degeneracy by an amount $\Delta'$. To first order in degenerate perturbation theory, $\Delta'$ is the smallest eigenvalue of the matrix $V_{\alpha,\beta} = \frac{1}{2}A(s)\braket{\alpha|\mathcal{H}_D|\beta}$ where $\ket{\alpha}$ and $\ket{\beta}$ denote specific degenerate eigenstates in the set of $N$. 

$V_{\alpha,\beta}$ is the negative adjacency matrix for states in this excited state manifold. For simplicity, we first assume that each state is connected to the same number of other states via single bit flips, i.e. the connectivity of these states is a regular graph. It is trivial, then, to check that the $N$ element vector

\begin{equation}
  \ket{v} = \frac{1}{\sqrt{N}} \begin{pmatrix}
    1  \\
    1  \\
    \vdots \\
    1
\end{pmatrix}
  \end{equation}
      
is an eigenvector of $V_{\alpha,\beta}$ with eigenvalue

\begin{equation}
  \lambda = - \frac{1}{N}\sum_{a=1}^N\sum_{b:(a,b)\in B} \tfrac{1}{2}A(s)
    \label{eqn:firstordercalc}
\end{equation}
where $B$ is the set of pairs of states $a,b$ connected by a single bit flip.

Because all elements $V_{\alpha,\beta} \leq 0$, this eigenvalue corresponds to the minimum eigenvalue~\cite{Perron1907}. Thus, $\Delta' = -\lambda$ and 

\begin{equation}
 E_1' = E_1 - \Delta'. 
 \end{equation}

For more complicated connection topologies of degenerate first excited states, that is, non-regular graphs, Eq.~\ref{eqn:firstordercalc} is still correct to first order and a good approximation as long as the connectivity graph is approximately regular.

\section{Using third eigengap}\label{app:eigengap}

Spectral analysis of quantum annealing typically considers the first eigengap \cite{Farhi2000,Farhi2001,Smelyanskiy2001}, since this value governs guarantees of adiabaticity in a closed system.  Since the 100 instances considered here have two antipodal classical ground states, this gap $E_1-E_0$ goes to zero at the end of the anneal, so analyzing its minimum is meaningless.  Instead we consider the third eigengap $E_3-E_0$.  The minimum third eigengap gives a consistent representation of the tunneling between ground-state and excited-state manifolds when the instantaneous ground state transitions from being mainly supported by classical excited states to being mainly supported by classical ground states.  Fig.~\ref{fig:exemplars} shows three exemplary instances under antimitigation, no mitigation, and mitigation, where the first and second gaps do not necessarily represent the desired tunneling event.

\end{document}